\documentclass[aps,prl,twocolumn,superscriptaddress]{revtex4}
\usepackage{graphics}
\begin{document}
\title{Growth control of GaAs nanowires using pulsed laser deposition with arsenic over pressure}
\author{X. W. Zhao, T. R. Lemberger, and F. Y. Yang}
\affiliation{Department of Physics, The Ohio State University, 191 W. Woodruff Ave. Columbus, OH 43210}

\date{\today}

\begin{abstract}
	Using pulsed laser ablation with arsenic over pressure, the growth conditions for GaAs nanowires have been systematically investigated and optimized.  Arsenic over pressure with As$_2$ molecules was introduced to the system by thermal decomposition of polycrystalline GaAs to control the stoichiometry and shape of the nanowires during growth.  GaAs nanowires exhibit a variety of geometries under varying arsenic over pressure, which can be understood by different growth processes via vapor-liquid-solid mechanism.  Single-crystal GaAs nanowires with uniform diameter, lengths over 20 $\mu$m, and thin surface oxide layer were obtained and can potentially be used for further electronic characterization.  

\end{abstract}


\maketitle

Semiconductor nanowires have attracted wide interest since the discovery of Si nanowire synthesis in 1998 \cite{Morales}.  A large variety of semiconductor nanowires and prototype nanowire electronic devices have been produced within past decade \cite{Huang,Patolsky}.  As one of the mostly studied semiconductors, GaAs nanowires have been synthesized by several techniques \cite{Haraguchi,Duan,Wu,Davidson,Yu}.  However, most of the works about GaAs nanowires focused on the synthesis \cite{Duan,Wu,Davidson,Yu,Duan2,Gudiksen,Roest,Harmand,Lee,Ihn,Skold,Borgstrom,Paiano,Guo}.  Very few reports discussed the electronic characteristics of GaAs nanowires, let alone GaAs nanowire devices \cite{Haraguchi}.  The lack of success in GaAs nanowire electronics is likely due to the difficulty of obtaining high-quality GaAs nanowires for device fabrication.  In this letter, we present an investigation of the growth mechanism of GaAs nanowires by pulsed laser deposition (PLD) with arsenic over pressure.  We have synthesized long, uniform single-crystal GaAs nanowires which are potentially suitable for electronic characterization and nanowire device fabrication.

	Most GaAs nanowires have been synthesized via vapor-liquid-solid (VLS) mechanism by pulsed laser deposition \cite{Duan,Duan2,Gudiksen,Roest}, molecular beam epitaxy (MBE) \cite{Wu,Harmand,Lee,Ihn}, or metal–organic vapor phase epitaxy (MOVPE) \cite{Haraguchi,Roest,Skold,Borgstrom,Paiano,Guo}.  MBE and MOVPE typically produce GaAs nanowires with tapered shape and length of 1 to 2 $\mu$m long, with a few exceptions up to $\sim$8 $\mu$m long.  The short length makes it challenging to fabricate electrical contacts on individual nanowires and the tapered shape complicates electronic characterization.  PLD has advantages in the synthesis of semiconductor nanowires with uniform diameter and length up to tens of $\mu$m.  However, it is well known that arsenic is highly volatile and tends to escape from GaAs surface at GaAs growth temperature.  To overcome this problem, arsenic over pressure with considerable As/Ga flux ratio is provided to grow stoichiometric GaAs in MBE and MOVPE.  Typical PLD systems do not have the capability to supply arsenic over pressure.  To our knowledge, there has been no report on controlling GaAs nanowire growth by PLD with arsenic over pressure.  In this work, we introduced an additional arsenic source in the PLD system in order to obtain stoichiometric growth of GaAs nanowires.  
	
	GaAs nanowires were fabricated using a PLD system comprised of a KrF excimer laser and a quartz tube furnace with base pressure below 5 mTorr.  A GaAs target was positioned at the upstream cool end of the furnace.  The laser energy density was maintained at 1.2 - 1.3 J/cm$^2$ on the target with a frequency of 10 Hz.  An Argon flow of 80 sccm was sent in before nanowire growth with the pressure controlled at 200 Torr.  Monodisperse Au nanoparticles with nominal diameter of 50 nm dispersed on thermally oxidized Si substrates were loaded in the quartz tube between the center and the downstream end of the furnace to obtain a growth temperature, $T_{\text{sub}}$, between 570 and 650$^{\circ}$C. The temperature at the center of the furnace, $T_{\text{center}}$, was controlled between 750 and 1000$^{\circ}$C. After the nanowire growth for 20 minutes, the furnace top was immediately opened and the quartz tube was cooled down quickly by a fan. Field-emission scanning electron microscopy (SEM) and high-resolution transmission electron microscopy (TEM), both equipped with energy dispersive spectrometer (EDS), were used to characterize the structure and composition of the GaAs nanowires.
	
	GaAs nanowires were first synthesized at $T_{\text{center}}$ = 750$^{\circ}$C and $T_{\text{sub}}$ from 570 to 650$^{\circ}$C $without$ arsenic over pressure.  The low resolution SEM image in Fig. 1a shows GaAs nanowires with length up to 100 $\mu$m.  However, a closer look at high magnifications reveals subtle details of the nanowires, as shown in Fig. 1b and 1c for nanowires grown at $T_{\text{sub}}$ = 640$^{\circ}$C and 610$^{\circ}$C, respectively.  Although some nanowires have uniform diameter of $\sim$50 nm and clean surface, the majority of the nanowires deviate from cylindrical wire geometry.  At $T_{\text{sub}}$ = 640$^{\circ}$C (Fig. 1b), many nanowires have rough surface and kinks with diameters considerably larger than 50 nm.  While at $T_{\text{sub}}$ = 610$^{\circ}$C (Fig. 1c), the nanowires are typically $\sim$50 nm in diameter, but with many `dust' like particles attached to them.  EDS analysis of the GaAs nanowires in Fig. 1b and 1c reveals more Ga than As.  

      We attribute the non-ideal geometries of the GaAs nanowires to the loss of arsenic from the GaAs nanowires during the growth.  Laser ablation on the GaAs target generates equal amounts of Ga and As vapors.  At growth temperature of 570 - 640$^{\circ}$C, arsenic in the just-grown GaAs nanowires tends to escape from the nanowires under equal Ga and As partial pressure, leaving Ga on the nanowire surface.  The metallic Ga on the nanowire surface serves as catalysts, just like the Au nanoparticles, and absorbs Ga and As vapors to initiate additional GaAs growth.  At $T_{\text{sub}}$ = 640$^{\circ}$C, metallic Ga remains as surface layer on the nanowires and catalyzes GaAs layer growth on the existing GaAs nanowire surface, resulting in much thicker GaAs nanowires.  As substrate temperature decreases to $T_{\text{sub}}$ = 610$^{\circ}$C, metallic Ga beads up to form nano-droplets on the GaAs nanowires and branches out additional GaAs growth, producing GaAs particles on the GaAs nanowires.  The TEM image of such a GaAs nanowire in Fig. 2a exhibits crystalline facets on these particles attached to the nanowire.  Electron diffraction on these particles confirms crystalline structure and the Ga/As ratio in these particles is slightly over 1, suggesting that these particles are mainly GaAs.  In order to synthesize stoichiometric GaAs nanowires with uniform diameter, arsenic over pressure is needed in the PLD synthesis process to compensate the loss of arsenic from the GaAs nanowires.

      Our first attempt used laser ablation on a mixture target of GaAs and arsenic powders to provide additional arsenic vapor.  However, the existence of arsenic powders in the target alters the effective laser energy density on the GaAs powders and significantly limits the GaAs nanowire growth.  Alternatively, we split the laser beam to ablate a GaAs target and a separate arsenic target simultaneously.  After arsenic atoms are ablated off from the target by the laser pulses, arsenic `smoke' forms in front of the targets.  The arsenic `smoke' severely scattered the laser beam and greatly reduced the generation of Ga and As vapors from the GaAs target, resulting in essentially no GaAs nanowire growth.

      A different approach to arsenic over pressure is to heat an arsenic source using the furnace.  Because it takes the furnace over an hour to reach the set temperature of 750 - 1000$^{\circ}$C, arsenic source cannot be loaded in the furnace before heating.  Otherwise, the slow heating process leads to uncontrollable arsenic vapor and contamination of the substrate.  In order to overcome this problem, we incorporated an in-situ transfer device to move the arsenic source into and out of the furnace without breaking the vacuum at high temperatures.  The arsenic source was initially positioned outside the furnace and inserted into the center of the furnace shortly before the nanowire growth.  
      
      We first used arsenic powders to generate arsenic vapor by sublimation at various furnace temperatures.  However, the side growth on GaAs nanowires persists, resulting in similar shapes as those shown in Fig. 1b and 1c.  Obviously, the arsenic vapor generated by sublimation did not compensate the loss of arsenic from the nanowire surface.  This is due to that the sublimation of arsenic produces As$_4$ molecules.  It is well known from the GaAs growth by MBE that As$_4$ molecules require fairly large energy to break into As atoms which are required to form GaAs.  Since As$_2$ molecules are much easier to break, most MBE systems use an arsenic cracker to break As$_4$ molecules into As$_2$ molecules in order to grow GaAs \cite{Harmand,Lee,Ihn}.  If an As$_2$ over pressure can be provided during GaAs nanowire growth, the abundant As$_2$ molecules cover the GaAs nanowire surface and can easily break into As atoms to compensate the lose of arsenic and maintain stoichiometric GaAs nanowires.  In the presence of As$_4$ over pressure, although As$_4$ molecules cover the nanowire surface, they cannot break into As atoms to compensate the lose of arsenic.
 
      Since arsenic cracker is not practical for PLD systems, we took a new route to generate As$_2$ vapor by thermal decomposition of polycrystalline GaAs, which releases As$_2$ molecules, using the in-situ transfer device described above \cite{Song}.  Fig. 1d shows the SEM image of GaAs nanowires grown at $T_{\text{center}}$ = 1000$^{\circ}$C and $T_{\text{sub}}$ = 640$^{\circ}$C with As$_2$ over pressure.  There are some large structures of a few $\mu$m on some nanowires.  EDS analysis on these large structures indicates a Ga/As ratio of 1.  These are GaAs crystals originated from epitaxial growth of GaAs on the side of some part of the nanowires in high As$_2$ over pressure at $T_{\text{center}}$ = 1000$^{\circ}$C, which should be distinguished from the Ga catalyzed growth as shown in Fig. 1b and 1c.  The epitaxial growth of GaAs on nanowires is similar to the deposition of GaAs films by MBE, which requires high As$_2$ over pressure to cover the GaAs surface with As for subsequent GaAs growth.  For nanowire growth, it is desirable to maintain an As$_2$ over pressure at a level that it just compensates the loss of As from the nanowires, but not enough to cover the nanowire surface with As for epitaxial growth. This optimal condition can be achieved by lowering the As$_2$ partial pressure at lower $T_{\text{center}}$.  The SEM image of GaAs nanowires grown at $T_{\text{center}}$ = 880$^{\circ}$C and $T_{\text{sub}}$ = 640$^{\circ}$C as shown in Fig. 1e exhibits long nanowires with uniform diameter and no side growth of GaAs.  However, they typically have zigzags because the substrate temperature is too high and the nanowires change growth directions.  
      
      At $T_{\text{sub}}$ = 570$^{\circ}$C and $T_{\text{center}}$ = 880$^{\circ}$C, we obtained straight GaAs nanowires with uniform diameters and length of over 20 $\mu$m, as shown in Fig. 1f.  The kinks in a small number of nanowires suggest that some defects still exist under this condition.  Further optimization should be able to reduce the defects in the nanowires.  A TEM image of one such GaAs nanowire is shown in Fig. 2b with a uniform diameter of 50 nm and smooth surface.  High resolution TEM image of a nanowire with diameter of 48 nm in Fig. 2c reveals single-crystal structure of the nanowire.  The fringes perpendicular to the wire axis correspond to the GaAs (111) planes with a spacing of 0.32 nm, indicating that the GaAs nanowires grow along $<$111$>$ direction.  The surface amorphous oxide layer is typically less than 1 nm thick due to the low level of O$_2$ in the quartz tube and the arsenic over pressure during the nanowire growth and the quick cool down after the growth.  
      
      We have systematically studied the synthesis conditions for GaAs nanowires using PLD.  Arsenic over pressure with As$_2$ molecules is introduced to the system by thermal decomposition of GaAs.  Long single-crystal GaAs nanowires with uniform diameter and thin surface oxide layer were obtained.  These attributes are desirable for further electronic characterization on GaAs nanowires.  
      


\begin{figure}
\centering
\caption{SEM images of GaAs nanowires (a) at low magnification, (b) grown at $T_{\text{center}}$ = 750$^{\circ}$C and $T_{\text{sub}}$ = 640$^{\circ}$C without As$_2$ over pressure, (c) grown at $T_{\text{center}}$ = 750$^{\circ}$C and $T_{\text{sub}}$ = 610$^{\circ}$C without As$_2$ over pressure, (d) grown at $T_{\text{center}}$ = 1000$^{\circ}$C and $T_{\text{sub}}$ = 640$^{\circ}$C with As$_2$ over pressure, (e) grown at $T_{\text{center}}$ = 880$^{\circ}$C and $T_{\text{sub}}$ = 640$^{\circ}$C with As$_2$ over pressure, and (f) grown at $T_{\text{center}}$ = 880$^{\circ}$C and $T_{\text{sub}}$ = 570$^{\circ}$C with As$_2$ over pressure (inset: a straight nanowire with uniform diameter).}
\label{1}
\end{figure}

\begin{figure}
\centering
\caption{TEM images of (a) a GaAs nanowire grown at $T_{\text{center}}$ = 750$^{\circ}$C and $T_{\text{sub}}$ = 610$^{\circ}$C without As$_2$ over pressure, (b) a GaAs nanowire grown at $T_{\text{center}}$ = 880$^{\circ}$C and $T_{\text{sub}}$ = 570$^{\circ}$C with As$_2$ over pressure (inset: the end of a nanowire with a Au cap), (c) a GaAs nanowire at high resolution grown at $T_{\text{center}}$ = 880$^{\circ}$C and $T_{\text{sub}}$ = 570$^{\circ}$C with As$_2$ over pressure.  The fine fringes in (c) are the GaAs (111) atomic planes with spacing of 0.32 nm.  The inset shows the nanowire with diameter of 48 nm.  The fringes in the inset have a separation four times of the GaAs (111) plane spacing, which is also visible in the main image.  These larger period fringes are the artifact of TEM.}
\label{2}
\end{figure}


\end{document}